\documentstyle[12pt]{article}
\begin{document}

\title{{\Large {\bf Computation of the conformal algebra of 1+3 decomposable
space-times}}}
\author{{Michael Tsamparlis, Dimitris Nikolopoulos and Pantelis S. Apostolopoulos} \\
{\small {\it Department of Physics, Section of
Astronomy-Astrophysics-Mechanics},}\\
{\small {\it University of Athens, Panepistemiopolis, Athens 157 83, GREECE}}}
\maketitle

\begin{abstract}
The conformal algebra of a 1+3 decomposable spacetime can be computed from
the Conformal Killing Vectors (CKV) of the 3-space. It is shown that the
general form of such a 3-CKV is the sum of a gradient CKV and a Killing or
homothetic 3-vector. It is proved that spaces of constant curvature always
admit such conformal Killing vectors. As an example the complete conformal
algebra of a G\"{o}del type spacetime is computed. Finally it is shown that
this method can be extended to compute the conformal algebra of more general
non-decomposable spacetimes.
\end{abstract}

PACS - numbers: 0420J.9530L

\section{Introduction}

A spacetime is called 1+3 decomposable if it admits a covariantly constant
non-null vector field. These spacetimes have a metric of the form:

\[
ds^2=\varepsilon (dx^1)^2+g_{\mu \nu }(x^\rho )dx^\mu dx^\nu 
\]
where $\mu ,\nu =2,3,4,$ $\partial /\partial x^1$ is the constant vector
field and $\varepsilon =$sign($\partial /\partial x^1).$ For $\varepsilon =1$
$(-1)$ we refer to 1+3 spacelike (timelike) spacetimes.

A complete classification of decomposable spacetimes in terms of the
holonomy group can be found in a recent paper by Capocci and Hall \cite{I1}.
The 1+3 spacetimes (timelike or spacelike) are only two of the possible 15
types of spacetimes resulting from this classification.

Although the 1+3 decomposable spacetimes are rather special they are
important because many well known spacetimes are of this form. For example
1+3 timelike spacetimes are all static spacetimes with vanishing vorticity
(which include many classical spacetimes). Concerning the 1+3 spacelike
spacetimes we refer the G\"{o}del-type spacetimes which have attracted
significant interest in recent years \cite{I2}. These spacetimes are
generalizations of the G\"{o}del spacetime.

Another less well known application of decomposable spacetimes is their
relation with the Affine Conformal Vectors (ACV). The existence of an ACV is
equivalent to the existence of a covariantly constant symmetric tensor field 
$K_{ab}$ which, as it has been shown by Hall and da Costa \cite{I3} leads to
a decomposable 1+3 spacetime or to a further 1+1+2 decomposition or,
finally, to a pp-wave spacetime which in general it is not decomposable.

Coley and Tupper \cite{I4} have studied in depth ACVs in spacetime and they
have shown that these vectors can be found from the Conformal Killing
Vectors (CKV) of a 1+3 decomposable spacetime or of a pp-wave spacetime.
They have found all 1+3 decomposable spacetimes which admit a CKV and they
have also given a general form for the CKV for the timelike and the
spacelike cases. Finally Hall and Capocci \cite{I1} have studied CKVs in
decomposable spacetimes in a different context and they have shown how their
results can be related to the general results of \cite{I4}.

Although both works referred above have done the major work on CKVs in 1+3
decomposable spacetimes, their methods are difficult to use in the
computation of the conformal algebra in a given 1+3 decomposable spacetime.
Indeed Coley and Tupper present the general solution in a special coordinate
system adapted to the conformal factor $\psi $ which is not known and
Capocci and Hall work on a more general level which is appropriate for
producing general results but difficult to work with in actual applications.
The purpose of this work is to move between these two approaches and present
a theory which will systematize the computation of the conformal algebra of
a 1+3 decomposable spacetime irrespectively of the coordinate system used.

In Section II we present the theory and show that the CKVs of the four
dimensional 1+3 spacetimes are computed from the CKVs of the invariant
3-spaces provided they are gradient vector fields in these spaces. We also
show that if the invariant 3-space is of constant curvature then there exist
always such gradient vector fields. The results are in complete agreement
with the results of Coley and Tupper \cite{I4}. In Section 3 we apply the
method and compute the complete conformal algebra of the RT (Rebou\c{c}%
as-Tiommo) spacetime \cite{I5} which is an interesting homogeneous spacetime
of the G\"{o}del-type with 7 KVs and no causal anomalies. Finally we show
how the method can be generalized to compute the algebra of certain
important non-decomposable spacetimes.

\section{The CKVs of a 1+3 decomposable spacetime}

\setcounter{equation}{0}

\noindent  Let {\it M} a 1+3 decomposable spacetime with metric $%
ds^2=\varepsilon (dx^1)^2+g_{\mu \nu }(x^\rho )dx^\mu dx^\nu $ ($\rho =2,3,4$%
). Let $\zeta ^a=\delta _1^a$ be the constant vector field defining the
preferred direction. The projection tensor $h_{ab}=g_{ab}-\varepsilon \zeta
_a\zeta _b$ is the metric of the 3-surface $x^1={const}$ (due to the obvious
isometry between these 3-spaces it is enough to work in one of them).
Covariant derivatives with respect to the metric $g_{ab}$ (respectively $%
h_{ab}$) shall be denoted with $";"$ (respectively $"|"$). The covariant
derivative of any smooth vector field $X^a$ on {\it M} can be decomposed
uniquely as follows:

\begin{equation}  \label{sx2.1}
X_{a;b}=\psi ({\bf X})g_{ab}+H_{ab}({\bf X})+F_{ab}({\bf X})
\end{equation}
where $H_{ab}({\bf X})$ is symmetric and traceless and $F_{ab}({\bf X}%
)=-F_{ba}({\bf X}).$

\noindent Obviously:

\begin{equation}  \label{sx2.2}
{\cal L}_Xg_{ab}=2\psi g_{ab}+2H_{ab}({\bf X}).
\end{equation}
Thus $X^a$ is a CKV iff $H_{ab}({\bf X})=0$. A similar decomposition holds
for the vectors $\xi _\alpha $ in the 3-space $x^1={const}.$:

\begin{equation}  \label{sx2.3}
\xi _{\alpha |\beta }=\lambda ({\bf \xi })g_{\alpha \beta }+{\cal H}_{\alpha
\beta }({\bf \xi })+{\cal F}_{\alpha \beta }({\bf \xi })
\end{equation}
and

\begin{equation}  \label{sx2.4}
{\cal L}_\xi g_{\alpha \beta }=2\lambda ({\bf \xi })g_{\alpha \beta }+2{\cal %
H}_{\alpha \beta }({\bf \xi })
\end{equation}
so that $\xi ^\alpha $ is a CKV of the 3-metric iff ${\cal H}_{\alpha \beta
}({\bf \xi })=0$.

\noindent 
In order to relate the CKVs of the 4-metric $g_{ab}$ with those of the
3-metric $g_{\alpha \beta }$ we decompose $X^a$ along and normally to $\zeta
^a$:

\begin{equation}  \label{sx2.5}
X_a=f(x^i)\zeta _a+X_a^{\prime }
\end{equation}
where $X_a^{\prime }=h_a^bX_b$. We define the vector $K_\alpha $ in the
3-space $x^1={const}$ by the requirement:

\[
X_a^{\prime }=K_\alpha \delta _a^\alpha 
\]
so that:

\begin{equation}
X_a=f(x^i)\zeta _a+K_\alpha \delta _a^\alpha .  \label{sx2.6}
\end{equation}
Using standard analysis we prove easily the following result:

\begin{equation}
X_{a;b}=f_{,b}\zeta _a+\varepsilon \stackrel{*}{K_\alpha }\delta _a^\alpha
\zeta _b+K_{\alpha |\beta }\delta _a^\alpha \delta _b^\beta  \label{sx2.7}
\end{equation}
where $","$ denotes partial derivative and an asterisk over a letter denotes
partial differentiation w.r.t. $x^1$ i.e. $\stackrel{*}{K_\alpha }=K_{\alpha
,1}$. Using the irreducible decompositions (\ref{sx2.1}) and (\ref{sx2.3})
and projecting equation (\ref{sx2.7}) along $\zeta ^a\zeta ^b,$ $\zeta
^ah_c^b,$ $h_c^a\zeta ^b,$ $h_c^ah_d^b$ we are able to express the
quantities $\psi ({\bf X}),H_{ab}({\bf X}),F_{ab}({\bf X})$ of $X^a$ in
terms of the corresponding quantities of the 3-vector $K_\alpha $ and the
partial derivatives of $f$. Using an obvious block matrix notation we have
the following result:

\begin{equation}
4\psi ({\bf X})=\stackrel{*}{f}+3\lambda ({\bf K})  \label{sx2.8}
\end{equation}

\begin{equation}
H_{ab}({\bf X})=\left( 
\begin{array}{cc}
\varepsilon (\stackrel{*}{f}-\psi ({\bf X})) & \frac \varepsilon 2(f_{,\mu
}+\varepsilon \stackrel{*}{K_\mu }) \\ 
\frac \varepsilon 2(f_{,\mu }+\varepsilon \stackrel{*}{K_\mu }) & {\cal H}%
_{\mu \nu }({\bf K})+\frac 14(\lambda -\stackrel{*}{f})g_{\mu \nu }
\end{array}
\right)  \label{sx2.9}
\end{equation}

\begin{equation}
F_{ab}({\bf X})=\left( 
\begin{array}{cc}
{\bf \bigcirc } & \frac \varepsilon 2(f_{,\mu }-\stackrel{*}{\varepsilon
K_\mu }) \\ 
-\frac \varepsilon 2(f_{,\mu }-\stackrel{*}{\varepsilon K_\mu }) & {\cal F}%
_{\mu \nu }({\bf K})
\end{array}
\right) .  \label{sx2.10}
\end{equation}
The condition that $X^a$ is a CKV, $H_{ab}({\bf X})=0,$ is equivalent to the
following equations:

\begin{equation}
\stackrel{\ast }{f}=\psi ({\bf X})  \label{sx2.11}
\end{equation}

\begin{equation}
f_{,\alpha }=-\varepsilon \stackrel{*}{K_\alpha }  \label{sx2.12}
\end{equation}

\begin{equation}
{\cal H}_{\alpha \beta }({\bf K})=0\qquad \mbox{and}\qquad \psi ({\bf X}%
)=\lambda ({\bf K}).  \label{sx2.13}
\end{equation}
Thus (\ref{sx2.11}) can be written:

\begin{equation}
\stackrel{\ast }{f}=\lambda ({\bf K}).  \label{sx2.14}
\end{equation}
Relations (\ref{sx2.11}) - (\ref{sx2.14}) have to be supplemented with the
integrability conditions of $f$ i.e. $\stackrel{*}{f}_{,\alpha }=(f_{,\alpha
})^{*}$ and $f_{,\alpha \beta }=f_{,\beta \alpha }$. Thus we obtain the
additional equations:

\begin{equation}
\lambda ({\bf K})_{,\alpha }=-\varepsilon \stackrel{**}{K}_\alpha
\label{sx2.15}
\end{equation}
\begin{equation}
\stackrel{\ast }{\cal F}_{\alpha \beta }({\bf K)}=0.  \label{sx2.16}
\end{equation}
Equations (\ref{sx2.12})-(\ref{sx2.16}) must be considered in two sets:
equations (\ref{sx2.12}), (\ref{sx2.14}) which calculate the function $%
f(x^i) $ in terms of $K_\alpha $ and equations (\ref{sx2.15})-(\ref{sx2.16})
which characterize the vector field $K_\alpha $ and in fact (as Coley and
Tupper have shown \cite{I4}) define completely the form of the 3-metric $%
g_{\alpha \beta }(x^\rho )$. Obviously $K_\alpha $ allows the computation of 
$X^a$ thus we concentrate on the relations (\ref{sx2.13}), (\ref{sx2.15}), (%
\ref{sx2.16}) concerning $K_\alpha $.

\noindent Equation (\ref{sx2.13}) means that $K^\alpha $ is a CKV of $%
g_{\alpha \beta }$ (already shown in \cite{I1}). Equation (\ref{sx2.15}) can
then be written:

\begin{equation}
\lambda ({\bf K})_{,\alpha }\mid _\beta =-\varepsilon \stackrel{**}{\lambda }%
({\bf K})g_{\alpha \beta }  \label{sx2.17}
\end{equation}
which is the fundamental equation (cf equation (A10) in the Appendix of \cite
{I4}) found by Coley and Tupper. The new result is equation (\ref{sx2.16})
which states that either the bivector ${\cal F}_{\alpha \beta }({\bf K)}$ is
independent of the coordinate $x^1$ or ${\cal F}_{\alpha \beta }({\bf K)}=0.$
We have the following Proposition.

\noindent {\bf Proposition 2.1.}

\noindent All proper CKVs $X^a$ of the four metric $g_{ab}$ are calculated
from the CKVs $K^\alpha $ of the 3-metric $g_{\alpha \beta }$ of the form:

\[
K^\alpha =\frac 1pm(x^1)\xi ^\alpha + L^\alpha (x^\rho ) 
\]
where

\noindent (a) $\xi _\alpha =A_{,\alpha }$ is a gradient CKVs of the 3-metric 
$g_{\alpha \beta }$ whose conformal factor satisfies the relation

\[
\lambda ({\bf \xi })_{|\alpha \beta }=p\lambda ({\bf \xi })g_{\alpha \beta } 
\]
\noindent (b) the function $m(x^1)$ satisfies equation (\ref{sx2.26})
(bellow).

\noindent (c) $L_\alpha $ is a KV or a HKV of the 3-metric $g_{\alpha \beta }
$ which is not a gradient vector field and its bivector equals $F_{ab}(K)$.
This implies that the non-gradient KVs of the 3-metric are identical with
those of the 4-metric.

\noindent {\bf Proof.}

\noindent Consider first the KVs and HKVs of the 3-metric $g_{\alpha \beta }$%
. These are defined by the condition $\lambda ({\bf K})_{,\alpha }=0$ which
by (\ref{sx2.15}) and (\ref{sx2.16}) implies that $K_\alpha (x^i)=K_\alpha
(x^\rho )$ otherwise the spacetime becomes \{1+1+2\} decomposable. We
conclude that the KVs of the 3-metric $g_{\alpha \beta }$ are KVs (recall
that $\psi ({\bf X})=\lambda ({\bf K})$) of the full metric $g_{ab}$. We
consider next the proper CKVs of the 3-metric. Due to the form of equation (%
\ref{sx2.17}) and the decomposability of spacetime, we are looking for
solutions of the form:

\begin{equation}
\lambda ({\bf K})=m(x^1)A(x^\rho )+B(x^\rho ).  \label{sx2.18}
\end{equation}
Differentiating and using equation (\ref{sx2.15}), we find:

\begin{equation}
m(x^1)A_{,\alpha }+B_{,\alpha }=-\varepsilon \stackrel{**}{K}_\alpha
\label{sx2.19}
\end{equation}
Differentiating again and using equation (\ref{sx2.17}), we get:

\begin{equation}
m(x^1)A_{|\alpha \beta }+B_{|\alpha \beta }=-\varepsilon \stackrel{**}{m}%
(x^1)A(x^\rho )g_{\alpha \beta }.  \label{sx2.20}
\end{equation}
Because $A(x^\rho ),B(x^\rho )$ are functions of $x^\rho $ only, equation (%
\ref{sx2.20}) implies that:

\begin{equation}  \label{sx2.21}
m(x^1)A_{|\alpha \beta }+\varepsilon \stackrel{**}{m}A(x^\rho )g_{\alpha
\beta }=C_1
\end{equation}

\begin{equation}
B_{|\alpha \beta }=-C_{1\alpha \beta }  \label{sx2.22}
\end{equation}
where $C_{1\alpha \beta }$ is a constant tensor. Integrating (\ref{sx2.19})
it follows:

\begin{equation}
\stackrel{\ast }{K}_\alpha =-\varepsilon \int m(x^1)dx^1A_{,\alpha
}-\varepsilon B_{,\alpha }x^1+D_\alpha (x^\rho ).  \label{sx2.23}
\end{equation}

\noindent Differentiating and taking the antisymmetric part we find that $%
D_\alpha (x^\rho )$ is a gradient vector field which we denote by $%
E_{,\alpha }(x^\rho ).$ Differentiating (\ref{sx2.21}) w.r.t. $x^1$ we find:

\begin{equation}  \label{sx2.24}
A_{|\alpha \beta }=pA(x^\rho )g_{\alpha \beta }
\end{equation}

\begin{equation}  \label{sx2.25}
\stackrel{\ast *}{m}+\varepsilon pm=C_2
\end{equation}

\noindent where $p,C_2$ are constants. Equation (\ref{sx2.24}) says that
when $p\neq 0$ then $A_{,\alpha }$ is a gradient CKV of $g_{\alpha \beta }$
with conformal factor $pA(x^\rho )$ and when $p=0$ then $A_{,\alpha }$ is a
gradient KV. Because as we shall see the gradient KVs do not interest us we
require $p\neq 0$. We compute easily $(A_{,\alpha }A^{,\alpha })_{|\beta
}=pA_{,\beta }^2$ thus $A_{,\alpha }$ is not a null vector field. Combining (%
\ref{sx2.24}) and (\ref{sx2.25}) with (\ref{sx2.21}) we find $C_{1\alpha
\beta }=0,C_2=0$. Thus $B_{,\alpha }$ is a gradient KV of the metric $%
g_{\alpha \beta }$ and the function $m(x^1)$ satisfies the equation:

\begin{equation}  \label{sx2.26}
\stackrel{\ast *}{m}+\varepsilon pm=0
\end{equation}

\noindent which means that ( $p\neq 0)$ $m(x^1)=\sin (\sqrt{|p|}x^1),\cos (%
\sqrt{|p|}x^1),\sinh (\sqrt{|p|}x^1),\cosh (\sqrt{|p|}x^1)$. Replacing these
results in (\ref{sx2.23}) and integrating we find:

\begin{equation}
K_\alpha =\frac 1pm(x^1)A_{,\alpha }-\frac \varepsilon 2B_{,\alpha
}(x^1)^2+E_{,\alpha }x^1+L_\alpha (x^\rho ).  \label{sx2.27}
\end{equation}
Differentiating (\ref{sx2.27}) and using the fact that $K_\alpha $ is a CKV
of $g_{\alpha \beta }$ with conformal factor $\lambda ({\bf K})$ we find:

\begin{equation}
E_{|\alpha \beta }=0  \label{sx2.28}
\end{equation}
\begin{equation}
L_{\alpha |\beta }=Bg_{\alpha \beta }+F_{\alpha \beta }(K).  \label{sx2.29}
\end{equation}

\noindent Equation (\ref{sx2.28}) means that the vector $E_{|\alpha }$ is
also a gradient KV of the three metric $g_{\alpha \beta }$ and equation (\ref
{sx2.29}) that the vector $L_\alpha $ is a Special CKV (or KV or HKV) of $%
g_{\alpha \beta }$ which is not a gradient vector field. Thus we conclude
that the only part of the proper CKV $K_\alpha $ which is a proper CKV, is
the gradient CKV $A_{,\alpha }$ of the metric $g_{\alpha \beta }$. Equation (%
\ref{sx2.27}) is the most general solution of the conditions (\ref{sx2.13}),
(\ref{sx2.15}), (\ref{sx2.16}).

\noindent However if one of the gradient KVs $B_{,\alpha }$ and $E,_\alpha $
is not equal to zero the spacetime decomposes further to a 1+1+2 spacetime
and Coley and Tupper \cite{I4} have shown that these spacetimes do not admit
any CKVs except in the trivial case they degenerate to Minkowski spacetime.
Thus these vectors must vanish and then the remaining vector $L_\alpha $ is
a KV or a HKV of the 3-metric whose bivector is equal to the bivector of $%
K_\alpha $.$\Box $\newline

\noindent This result is fundamental and allows us to compute the conformal
algebra of the four metric $g_{ab}$. We note that this result is consistent
with equations (2.6) and (3.6) of \cite{I4}.

\noindent Coley and Tupper have used equation (\ref{sx2.17}) to calculate
all possible 1+3 spacetimes. We shall use (\ref{sx2.15}) and (\ref{sx2.16})
to calculate the CKVs in any 1+3 spacetime.

\noindent The conformal factor $\lambda ({\bf K})$ of the CKV $K^\alpha $
equals:

\begin{equation}
\lambda ({\bf K})=m(x^1)\lambda ({\bf \xi })+b  \label{sx2.36}
\end{equation}
where $b(=$0 or 1$)$ is the conformal factor of the vector $L_\alpha $.
Taking into account these results the function $f(x^a)$ is computed from
equations (\ref{sx2.12}),(\ref{sx2.14}) as follows:

\begin{equation}
f(x^a)=-\frac \varepsilon p\stackrel{*}{m}(x^1)\lambda ({\bf \xi })+bx^1.
\label{sx2.38}
\end{equation}
Finally it is useful to note how one can use $\xi _\alpha $ and determine
all 1+3 metrics in a simple manner. Indeed equation (\ref{sx2.24}) states
that $A_{,\alpha }(x^\rho )$ is a (non-null) gradient CKV of the 3-metric $%
g_{\alpha \beta }$ hence Petrov's result \cite{I6} quoted in \cite{I4}
applies and means that there exist coordinates ($x^2,x^A$) ($A,B,C=3,4$) in
which:

(i) $A_{,\alpha }=e\delta _\alpha ^2$, i.e. $A(x^2)=$ $ex^2+C$ where $e$ is
the sign of $A_{,\alpha }$

(ii) The 3-metric $g_{\alpha \beta }$ is written as:

\begin{equation}
g_{\alpha \beta }=eg_{22}(x^2)(dx^2)^2+\frac
1{g_{22}(x^2)}p_{AB}(x^C)dx^Adx^B  \label{sx2.40}
\end{equation}
where ($p=$constant):

\begin{equation}
g_{22}(x^2)=\frac 1{2p\int A(x^2)dx^2}  \label{sx2.42}
\end{equation}
where (without loss of generality) we take $g_{22}>0$. Defining the new
coordinate $x^{2^{\prime }}$ by the equation $\sqrt{g_{22}(x^2)}%
dx^2=dx^{2^{\prime }}$ the metric $g_{\alpha \beta }$ is written:

\begin{equation}  \label{sx2.44}
g_{\alpha \beta }=e(dx^{2^{\prime }})^2+M^2(x^{2^{\prime
}})p_{AB}(x^C)dx^Adx^B
\end{equation}
where

\begin{equation}
M^2(x^{2^{\prime }})=1/g_{22}=2p\int A(x^2)dx^2.  \label{sx2.46}
\end{equation}
Differentiating twice (\ref{sx2.46}) w.r.t. $x^{2^{\prime }}$ and using the
definition of the coordinate $x^{2^{\prime }}$ one shows that $%
M^2(x^{2^{\prime }})$ satisfies the equation

\[
\frac{d^2M}{d^2x^{2^{\prime }}}-epM=0. 
\]
Equation (\ref{sx2.44}) coincides formally with equation (A19) of Coley and
Tupper \cite{I4} but now the coefficient $M(x^{2^{\prime }})$ has a direct
geometrical meaning because according to (\ref{sx2.46}) it is related to the
conformal factor of the CKV of the 3-metric $g_{\alpha \beta }$. Furthermore
depending on the value of the product $ep$ it has solutions $\sin ,$ $\cos ,$
$\sinh ,$ $\cosh ,$ $ax^{2^{\prime }}+b$ ($a,b$ constants). We note that the
case $p=0$ corresponds to SCKVs of the 3-metric $g_{\alpha \beta }$ \cite{I7}%
.\newline

\section{The case of spaces of constant curvature}

\setcounter{equation}{0}

The fact that the CKVs of the 3-metric which produce the CKVs of the
4-metric must be of the form $K^\alpha =\frac 1pm(x^1)\xi ^\alpha +L^\alpha
(x^\rho )$ is a strong condition and one wonders that perhaps there are very
few 3-dimensional metrics which admit such vector fields. We show the
following result.

\noindent {\bf Proposition 3.1.}

\noindent The metrics of spaces of constant curvature of dimension $n$ admit 
$n+1$ gradient CKVs.

\noindent {\bf Proof}

\noindent Consider two metrics $g_{ij}$ and $\bar{g}_{ij}$ which are
conformally related:

\begin{equation}
g_{ij}=N^2(x^k)\bar{g}_{ij}.  \label{sx3.1}
\end{equation}
Taking the Lie derivative of (\ref{sx3.1}) w.r.t. the vector field $X^a$ and
using (\ref{sx2.1}) and (\ref{sx2.2}), we find:

\begin{equation}  \label{sx3.2}
\psi ({\bf X})={\bf X}(\ln N)+\phi ({\bf X})
\end{equation}

\begin{equation}
H_{ij}({\bf X})=N^2\bar{H}_{ij}({\bf X})  \label{sx3.3}
\end{equation}
\begin{equation}
F_{ij}({\bf X})=N^2\bar{F}_{ij}({\bf X})-2NN_{[,i}\overline{X}_{j]}.
\label{sx3.4}
\end{equation}
We specialize the metric $\bar{g}_{ij}$ to be the flat metric (with
Euclidean or Lorentzian character) $\eta _{ij}$ so that $g_{ij}$ is
conformally flat. Then ${\bf X}$ is a vector of the flat conformal algebra
whose in generic form in Cartesian coordinates is \cite{I8}\cite{I9}:

\begin{equation}  \label{sx3.5}
X^i=a^i+a_{..j}^ix^j+bx^i+2({\bf b\cdot x})x^i-b^i({\bf x\cdot x})
\end{equation}
where $a^i,$ $a_{..j}^i,$ $b,$ $b^i$ are constants and $({\bf x\cdot x}%
)=\eta _{ij}x^ix^j$. KVs are defined by the constants $a^i,$ $a_{..j}^i$,
there exists only one HKV defined by the constant $b$ and the constants $b^i$
define the remaining four Special CKVs. \newline

\noindent Using (\ref{sx3.5}) we compute for the generic vector field:

\begin{equation}
F_{ij}({\bf X})=X_{[i,j]}=(\overline{a}_{ij}+4\overline{b}%
_{[j}x_{i]})N^2+2(\ln N)_{[,j}X_{i]}.  \label{sx3.6}
\end{equation}
Assume now that the metric $g_{ij}$ is a metric of constant curvature. Then
the conformal factor:

\[
N(x^i)=\frac 1{1+\frac K4({\bf x\cdot x})} 
\]
where $K=\frac R{n(n-1)}$ and $R$ is the Gaussian curvature of space.
Introducing $N(x^i)$ in (\ref{sx3.6}) we compute:

\begin{equation}
F_{ij}({\bf X})=a_{ij}-KNx_{[j}a_{i]}-KNx_{[j}a_{i]r}x^r+4N\bar{b}%
_{[j}x_{i]}.  \label{sx3.7}
\end{equation}
Following the notation of \cite{I9} we write the CKVs of the flat space as
follows:

\[
{\bf P}_i=\partial _i,\qquad {\bf M}_{ij}=x_i\partial _j-x_j\partial
_i\qquad \qquad (\mbox{KVs}) 
\]
\[
{\bf H}=x^i\partial _i\qquad \qquad \qquad \qquad \qquad \qquad \quad (%
\mbox{HKV}) 
\]
\[
{\bf K}_i=2x_i{\bf H}-({\bf x\cdot x}){\bf P}_i\qquad \qquad \qquad \qquad (%
\mbox{SCKVs}). 
\]
Then (\ref{sx3.7}) implies:

\begin{equation}
F_{ij}({\bf P}_r)=-KNx_{[j}\delta _{i]}^r  \label{sx3.8}
\end{equation}
\begin{equation}
F_{ij}({\bf M}_{rs})=a_{ij}({\bf M}_{rs})-KNx_{[j}a_{i]k}({\bf M}_{rs})x^k
\label{sx3.9}
\end{equation}
\begin{equation}
F_{ij}({\bf H})=0  \label{sx3.10}
\end{equation}
\begin{equation}
F_{ij}({\bf K}_r)=4Nx_{[i}\delta _{j]}^r.  \label{sx3.11}
\end{equation}
We note that the only vector whose bivector vanishes identically (hence it
is a gradient vector field) is the proper CKV ${\bf H}$. However we also
note that the bivectors of the proper CKVs ${\bf P}_i-\frac K4{\bf K}_i$
also vanish. Thus we have proved that in a n-dimensional space (of Euclidean
or Lorentzian character) of constant curvature there exist always $(n+1)$
gradient CKVs. These vectors can be used to compute the conformal algebra of
any 1+3 spacetime whose 3-d subspaces are spaces of constant curvature. $%
\Box $\newline
\newline
\newline
\noindent The above result is important because: \newline

\noindent a) It says that there are at most four (non flat) spacetimes whose
3-d subspaces are spaces of constant curvature. Two spacetimes are of the
1+3 timelike type and are the Einstein spacetime ($K<0$) and anti-Einstein
spacetime ($K>0$). The other two spacetimes are of the 1+3 spacelike type
and have been found by Rebou\c{c}as and Tiommo (RT metric) \cite{I5} and
Rebou\c{c}as and Texeira (ART metric) \cite{I10}. The RT spacetime is of
G\"{o}del type and has constant curvature $R<0$ whereas the ART spacetime is
not of Godel type and has $R>0.$ Both spacetimes can be found from the
general types described in \cite{I4} (Note that in \cite{I4} only the
spacetimes which satisfy the strong energy condition have been considered).

\noindent b) The four space--times mentioned above have 7 KVs a fact that
makes them interesting and very rare. These KVs have been computed in Refs 
\cite{I5},\cite{I10}. However the computation is not necessary and one can
arrive at this result without any explicit computation. Indeed we have shown
that the KVs and the HKVs of the 3-space are also KVs and HKVs of the full
1+3 spacetime. Because there are no HKVs we have $\frac{3\cdot 4}2=6$ KVs
for the 3-spaces of constant curvature. If we include in these the constant
vector field decomposing the spacetime (i.e. $\partial /\partial x^1$) the
total number of KVs becomes seven. \newline

\noindent c) All these spacetimes are conformally flat. This can be
established either by showing directly that their Weyl tensor vanishes or by
showing that they admit a Conformal Algebra of 15 CKVs. The first method is
straighforward once one has the metric. The second method is interesting
because it does not require a knowledge of the metric. Indeed we have shown
that in a three dimensional space there are four proper CKVs (the ${\bf H}$, 
${\bf P}_i-\frac K4{\bf K}_i$) with vanishing bivector. From (\ref{sx2.24}),(%
\ref{sx2.26}) we note that for each gradient CKV we obtain two functions $%
m(x^1)$ (depending on the sign of $p$) thus two CKVs of spacetime. Hence the
four CKVs of the 3-metric produce eight CKVs of the 1+3 metric which implies
that its algebra consists of 15 CKVs, hence spacetime is conformally flat. 
\newline
\newline
\noindent Finally using (\ref{sx3.2}) it is easy to prove that the conformal
factor of the generic CKV $X^i$ in (\ref{sx3.5}) is given by the following
formula:

\begin{equation}
\psi ({\bf K})=-\frac 12KNa^r\bar{x}_r+(2N-1)b+2N({\bf b\cdot x}).
\label{sx3.12}
\end{equation}
From (\ref{sx3.12}) we conclude that the vectors, ${\bf M}_{ij}$ are KVs
whereas the vectors ${\bf P}_i,$ ${\bf H}$ and ${\bf K}_i$ are proper CKVs
of the 3-metric $g_{\alpha \beta }$.

\section{Application}

\setcounter{equation}{0}

We apply the results of Section 3 to compute the CKVs of the RT spacetime 
\cite{I5} whose metric is:

\begin{equation}  \label{sx4.1}
ds^2=-\left[ dt+H(r)d\Phi \right] ^2+dr^2+D^2(r)d\Phi ^2+dz^2
\end{equation}
where

\begin{equation}
H(r)=-\frac 1\Omega \sinh ^2(\Omega r),\qquad \qquad D(r)=\frac 1{2\Omega
}\sinh (2\Omega r).  \label{sx4.2}
\end{equation}
As we have already mentioned the KVs of this metric have been computed in 
\cite{I5} and there remain the eight proper CKVs to be found for the
completion of the conformal algebra. It is easy to show that the 3D RT
spacetime z=constant is a space of constant curvature with $R=-6\Omega ^2$,
hence it is possible to apply the results of the previous section and
compute the CKVs of the full RT spacetime.\newline

\noindent The method we follow in the computation is the following: The 3-d
RT metric $ds_{RT}^2$ has the same CKVs with the 3-dimensional Lorentzian
space -$d\tau ^2+dx^2+dy^2$. Thus we need the CKVs of the 3-d Lorentz metric
whose bivector vanishes. These vectors are the vectors ${\bf H}$, ${\bf P}%
_i-\frac K4{\bf K}_i$ which are known in the Lorentz coordinates $\{\tau
,x,y\}$ but not in the coordinates $\{t,r,\Phi \}$. If we find the
transformation equations:

\[
\tau (t,r,\Phi ),\qquad x(t,r,\Phi ),\qquad y(t,r,\Phi ) 
\]
then we shall be able to express these vectors in the new coordinates and
have the required CKVs. In fact it is enough to find only one proper CKV
which is a gradient vector and then compute the rest three by taking the
commutators of this vector with the seven KVs already known. Obviously the
convenient vector to use is the proper CKV ${\bf H.}$

\noindent Because the 3-dimensional RT space is of constant curvature $%
R=-6\Omega ^2$ we have $K=-\Omega ^2$ and:

\begin{equation}
ds_{RT}^2=\frac 1{\Omega ^2\left[ 1-\frac 14\left( x^2+y^2-\tau ^2\right)
\right] ^2}ds_L^2  \label{sx4.3}
\end{equation}
where $ds_L^2=-d\tau ^2+dx^2+dy^2$. Using the transformation:

\[
\Phi =\phi -\Omega t,\qquad \qquad \Omega =\frac 1a 
\]

\noindent we get:

\begin{equation}  \label{sx4.4}
ds_{RT}^2=-\cosh ^2(\frac ra)dt^2+dr^2+a^2\sinh ^2(\frac ra)d\phi ^2.
\end{equation}
After enough trial and error we have managed to show that the transformation 
$(t,r,\phi )\longrightarrow (\tau ,x,y)$ which makes (\ref{sx4.3}) valid is (%
$0<\phi <2\pi ,0<t<\infty $):

:

\begin{equation}
\begin{array}{c}
\tau =-\frac{2\sin (\frac ta)\cosh (\frac ra)}{1-\cosh (\frac ra)\cos (\frac
ta)} \\ 
\\ 
x=\frac{2\sinh (\frac ra)\cos \phi }{1-\cosh (\frac ra)\cos (\frac ta)} \\ 
\\ 
y=\frac{2\sinh (\frac ra)\sin \phi }{1-\cosh (\frac ra)\cos (\frac ta)}
\end{array}
\label{sx4.5}
\end{equation}
In these coordinates the conformal factor $\frac 1{\left[ 1-\frac 14\left(
x^2+y^2-\tau ^2\right) \right] ^2}$ becomes:

\begin{equation}
N(r,t)=\frac 12\left[ 1-\cosh (\frac ra)\cos (\frac ta)\right] .
\label{sx4.6}
\end{equation}
Using the transformation (\ref{sx4.5}) we find that in the coordinates $%
(t,r,\phi )$ the vector ${\bf H}\equiv {\bf \xi }_1$ is:

\begin{equation}
{\bf \xi }_1=-a\frac{\sin (\frac ta)}{\cosh (\frac ra)}\partial _t-a\cos
(\frac ta)\sinh (\frac ra)\partial _r.  \label{sx4.7}
\end{equation}
The conformal factor of ${\bf \xi }_1$ is read from (\ref{sx3.12}):

\begin{equation}
\lambda ({\bf \xi }_1)=2N-1=-\cosh (\frac ra)\cos (\frac ta).  \label{sx4.8}
\end{equation}
Following the same procedure we recover the six KVs of the 3-dimensional RT
metric (these are the three vectors ${\bf M}_{AB}$ and the three vectors $%
{\bf P}_A-\frac 14{\bf K}_A$ $(K=-1)$ where $A,B=\tau ,x,y$) which have been
already computed in \cite{I5}. To find the rest three gradient proper CKVs
of the 3-dimensional RT metric we take the Lie bracket of the proper CKV $%
{\bf H}$ with these KVs. We find:

\begin{equation}
{\bf \xi }_2=-a\frac{\cos (\frac ta)}{\cosh (\frac ra)}\partial _t+a\sin
(\frac ta)\sinh (\frac ra)\partial _r  \label{sx4.9}
\end{equation}
\begin{equation}
\lambda ({\bf \xi }_2)=\cosh (\frac ra)\sin (\frac ta)  \label{sx4.10}
\end{equation}
\begin{equation}
{\bf \xi }_3=-a\cosh (\frac ra)\cos (\phi )\partial _r+\frac{\sin (\phi )}{%
\sinh (\frac ra)}\partial _\phi  \label{sx4.11}
\end{equation}
\begin{equation}
\lambda ({\bf \xi }_3)=-\sinh (\frac ra)\cos (\phi )  \label{sx4.12}
\end{equation}
\begin{equation}
{\bf \xi }_4=a\cosh (\frac ra)\sin (\phi )\partial _r+\frac{\cos (\phi )}{%
\sinh (\frac ra)}\partial _\phi  \label{sx4.13}
\end{equation}
\begin{equation}
\lambda ({\bf \xi }_4)=\sinh (\frac ra)\sin (\phi ).  \label{sx4.14}
\end{equation}
For each conformal factor we calculate the constant $p$ by using equation (%
\ref{sx2.24}) and subsequently the function $m(z)$ using (\ref{sx2.26}).
Finally $f(x^i)$ is computed from equation (\ref{sx2.38}) and we write the
CKVs of the 4-metric using the formula:

\[
X^a=f(x^i)\zeta ^a+m(z)\xi ^a. 
\]
The conformal factor of this vector is $(B=0)$:

\[
\psi ({\bf X})=m(z)\lambda ({\bf \xi }). 
\]
The results of these calculations are the following eight proper CKVs of the
RT spacetime ($k=1,2,3,4$ and $\alpha =0,1,2$):

\[
X_{(k)\alpha }=a^2A_{k,\alpha \mbox{ },}X_{(k)3}=-a^2A_{k,3}. 
\]
with conformal factors:

\[
\psi _k=A_k 
\]
and

\[
X_{(k+4)\alpha }=a^2B_{k,\alpha \mbox{ ,}}X_{(k+4)3}=-a^2B_{k,3}. 
\]
with conformal factors:

\[
\psi _{k+4}=B_k 
\]
where

\[
A_k=\cosh (\frac ra)\left[ \cos (\frac ta)\cos (\frac za),\sin (\frac
ta)\cos (\frac za),\cos (\frac ta)\sin (\frac za),\sin (\frac ta)\sin (\frac
za)\right] 
\]

\[
B_k=\sinh (\frac ra)\left[ \sin \phi \cos (\frac za),\cos \phi \cos (\frac
za),\cos \phi \sin (\frac za),\sin \phi \sin (\frac za)\right] . 
\]

\noindent To write the conformal algebra of RT spacetime in a compact form
it is necessary to define a new basis of CKVs. For the Killing Vectors we
define:

\[
\begin{array}{c}
\begin{array}{cc}
C_0=K_1+K_3\qquad \qquad & M_{12}=K_1-K_3 \\ 
C_1=K_7-K_4\qquad \qquad & M_{10}=K_5+K_6 \\ 
C_2=K_5-K_6\qquad \qquad & M_{20}=K_4+K_7
\end{array}
\\ 
\\ 
K_2=a\partial _z
\end{array}
\]
where:

\[
\begin{array}{c}
K_1+K_3=a\partial _t\qquad \qquad K_1-K_3=\partial _\phi \\ 
\\ 
K_5-K_6=a\tanh (\frac ra)\sin \phi \sin (\frac ta)\partial _t-a\sin \phi
\cos (\frac ta)\partial _r-\coth (\frac ra)\cos \phi \cos (\frac ta)\partial
_\phi \\ 
\\ 
K_7-K_4=a\tanh (\frac ra)\cos \phi \sin (\frac ta)\partial _t-a\cos \phi
\cos (\frac ta)\partial _r+\coth (\frac ra)\sin \phi \cos (\frac ta)\partial
_\phi \\ 
\\ 
K_5+K_6=a\tanh (\frac ra)\cos \phi \cos (\frac ta)\partial _t+a\cos \phi
\sin (\frac ta)\partial _r-\coth (\frac ra)\sin \phi \sin (\frac ta)\partial
_\phi \\ 
\\ 
K_4+K_7=a\tanh (\frac ra)\sin \phi \cos (\frac ta)\partial _t+a\sin \phi
\sin (\frac ta)\partial _r+\coth (\frac ra)\cos \phi \sin (\frac ta)\partial
_\phi
\end{array}
\]

\noindent For the CKVs we introduce the following notation:

\[
\begin{array}{cc}
I_0=X_2\qquad \qquad & Z_0=X_4 \\ 
I_1=X_6\qquad \qquad & Z_1=X_7 \\ 
I_2=X_5\qquad \qquad & Z_2=X_8
\end{array}
\]
and:

\[
H_1=X_1\qquad \qquad H_2=X_3. 
\]
Then the conformal algebra is written as follows:

\[
\left[ M_{\alpha \beta },H_1\right] =\left[ M_{\alpha \beta },H_2\right]
=\left[ M_{\alpha \beta },K_2\right] =\left[ C_\alpha ,K_2\right] =0 
\]

\[
\left[ M_{\alpha \beta },I_\gamma \right] =\eta _{\alpha \gamma }I_\beta
-\eta _{\beta \gamma }I_\alpha \qquad \qquad \left[ M_{\alpha \beta
},Z_\gamma \right] =\eta _{\beta \gamma }Z_\alpha -\eta _{\alpha \gamma
}Z_\beta 
\]

\[
\left[ I_\alpha ,C_\beta \right] =\eta _{a\beta }H_1\qquad \qquad \left[
Z_\alpha ,C_\beta \right] =\eta _{a\beta }H_2 
\]

\[
\left[ Z_a,I_\beta \right] =\eta _{a\beta }K_2\qquad \qquad \left[
H_1,H_2\right] =K_2 
\]

\[
\left[ H_1,K_2\right] =\frac{H_2}2\qquad \qquad \left[ K_2,H_2\right] =\frac{%
H_1}2 
\]

\[
\left[ I_a,K_2\right] =\eta _{a\beta }Z_\beta \qquad \qquad \qquad \qquad
\left[ K_2,Z_a\right] =\eta _{a\beta }I_\beta 
\]

\[
\left[ M_{\alpha \beta },C_\gamma \right] =\eta _{\beta \gamma }C_\alpha
-\eta _{\alpha \gamma }C_\beta 
\]

\[
\left[ M_{\alpha \beta },M_{\gamma \delta }\right] =\eta _{\alpha \delta
}M_{\beta \gamma }+\eta _{\beta \gamma }M_{\alpha \delta }+\eta _{\alpha
\gamma }M_{\beta \delta }+\eta _{\beta \delta }M_{\alpha \gamma } 
\]
where $\eta _{\alpha \beta }=diag(-1,1,1)$ and $\alpha ,\beta ,\gamma
,\delta =0,1,2$ plus the three relations:

\[
\left[ M_{12},I_1\right] =-I_2\qquad \qquad \left[ M_{12},I_2\right]
=I_1\qquad \qquad \left[ Z_0,C_0\right] =H_2 
\]

\section{Conclusions}

\noindent We have shown that it is possible to develop a systematic method
for the computation of the conformal algebra of any 1+3 decomposable
spacetime. The key step of the method, in the special case of 3-d metrics of
constant curvature, is to find a coordinate transformation between the
coordinates in which the metric is given and the Lorentz coordinates of 3-d
Minkowski spacetime. This transformation is not trivial but it is possible
to be found by a trial and error procedure.. The method we have developed is
systematic and exhibits the geometrical character of the various crucial
quantities involved and, futhermore, it lays down the ground for the
systematic computation of higher collineations in these spacetimes.

\noindent One can extent this method to calculate the conformal algebra of
non-decomposable spacetimes which are conformally related to 1+3
decomposable spacetimes. For example let us consider the important case of
Friedman-Robertson-Walker (FRW) spacetimes. In conformal coordinates the
line element of these spaces can be written $ds^2=R^2(\tau )[-d\tau
^2+d\sigma ^2]$ where $\tau $ is the proper time along the comoving
observers, $R(\tau )$ is the scale factor and $d\sigma ^2$ is the metric of
a 3-d space of constant curvature. Thus essentially the FRW spacetimes are
conformally related to a 1+3 decomposable spacetimes whose 3-d invariant
subspaces are spaces of constant curvature. Having shown that we can always
compute the CKVs of these spaces and taking into account that the algebra of
conformally related spaces is the same we compute easily and systematically
the CKVs of the FRW spacetimes in the coordinates used. These vectors have
been computed in \cite{I9} and there is no point to repeat the computation
here. Instead using a similar procedure we give in the Appendix the CKVs of
the anti-DeSitter spacetime which, to our knowledge, they have not been
published before.

\section{Acknowledgments}

\noindent PSA is supported by Foundation of National Scholarships of Greece
(I.K.Y.) through a Postgraduate Bursary.

\section{Appendix}

The line element of anti-DeSitter spacetime is:

\[
ds^2=-\cosh ^2(\frac ra)dt^2+dr^2+a^2\sinh ^2(\frac ra)(d\theta ^2+\sin
^2\theta d\phi ^2). 
\]

\noindent The proper CKVs of this spacetime are the following:

\[
\begin{array}{c}
{\bf \xi }_1=a\frac{\sin (\frac ta)}{\cosh (\frac ra)}\partial _t+a\cos
(\frac ta)\sinh (\frac ra)\partial _r \\ 
\\ 
\psi _1=\cos (\frac ta)\cosh (\frac ra)
\end{array}
\]

\[
\begin{array}{c}
{\bf \xi }_2=-a\frac{\cos (\frac ta)}{\cosh (\frac ra)}\partial _t+a\sin
(\frac ta)\sinh (\frac ra)\partial _r \\ 
\\ 
\psi _2=\sin (\frac ta)\cosh (\frac ra)
\end{array}
\]

\[
\begin{array}{c}
{\bf \xi }_3=a\cosh (\frac ra)\sin \theta \cos \phi \partial _r-\frac{\sin
\phi }{\sinh (\frac ra)\sin \theta }\partial _\phi +\frac{\cos \theta \cos
\phi }{\sinh (\frac ra)}\partial _\theta \\ 
\\ 
\psi _3=\sinh (\frac ra)\sin \theta \cos \phi
\end{array}
\]

\[
\begin{array}{c}
{\bf \xi }_4=a\cosh (\frac ra)\sin \theta \sin \phi \partial _r+\frac{\cos
\phi }{\sinh (\frac ra)\sin \theta }\partial _\phi +\frac{\cos \theta \sin
\phi }{\sinh (\frac ra)}\partial _\theta \\ 
\\ 
\psi _4=\sinh (\frac ra)\sin \theta \sin \phi
\end{array}
\]

\[
\begin{array}{c}
{\bf \xi }_5=a\cosh (\frac ra)\cos \theta \partial _r-\frac{\sin \theta }{%
\sinh (\frac ra)}\partial _\theta \\ 
\\ 
\psi _5=\sinh (\frac ra)\cos \theta
\end{array}
\]

\end{document}